\documentclass[preprint]{aastex62}
\usepackage[utf8]{inputenc}
\usepackage{color}
\usepackage{soul}
\usepackage{xcolor}

\usepackage{amsmath,amsfonts,amssymb,amsthm, bm}

\begin{document}

\title{\LARGE{\textbf{Magnetohydrostatic Modeling of the Solar Atmosphere}}}

\correspondingauthor{Xiaoshuai Zhu} \email{xszhu@bao.ac.cn}

\author{Xiaoshuai Zhu}
\affil{State Key Laboratory of Solar Activity and Space Weather, National Space Science Center, Chinese Academy of Sciences, Beijing, China}
\affiliation{National Astronomical Observatories, Chinese Academy of Sciences, Beijing, China}

\author{Thomas Neukirch}
\affiliation{School of Mathematics and Statistics, University of St Andrews, St Andrews, KY16 9SS, UK}

\author{Thomas Wiegelmann}
\affiliation{Max-Planck-Institut f\"ur Sonnensystemforschung \\
Justus-von-Liebig-Weg 3, D-37077 G\"ottingen, Germany}


\begin{abstract}
Understanding structures and evolutions of the magnetic fields and plasma in multiple layers on the Sun is very important. A force-free magnetic field which is an accurate approximation of the solar corona due to the low plasma $\beta$ has been widely studied and used to model the coronal magnetic structure. While the force-freeness assumption is well satisfied in the solar corona, the lower atmosphere is not force-free given the high plasma $\beta$. Therefore, a magnetohydrostatic (MHS) equilibrium which takes into account plasma forces, such as pressure gradient and gravitational force, is considered to be more appropriate to describe the lower atmosphere. This paper reviews both analytical and numerical extrapolation methods based on the MHS assumption for calculating the magnetic fields and plasma in the solar atmosphere from measured magnetograms.
\end{abstract}

\keywords{Sun: magnetic field, Sun: photosphere, Sun: chromosphere, Sun: corona}

\section{Introduction}\label{se:intro}

%

\begin{figure*}[t]
\centering
\includegraphics[width=13cm]{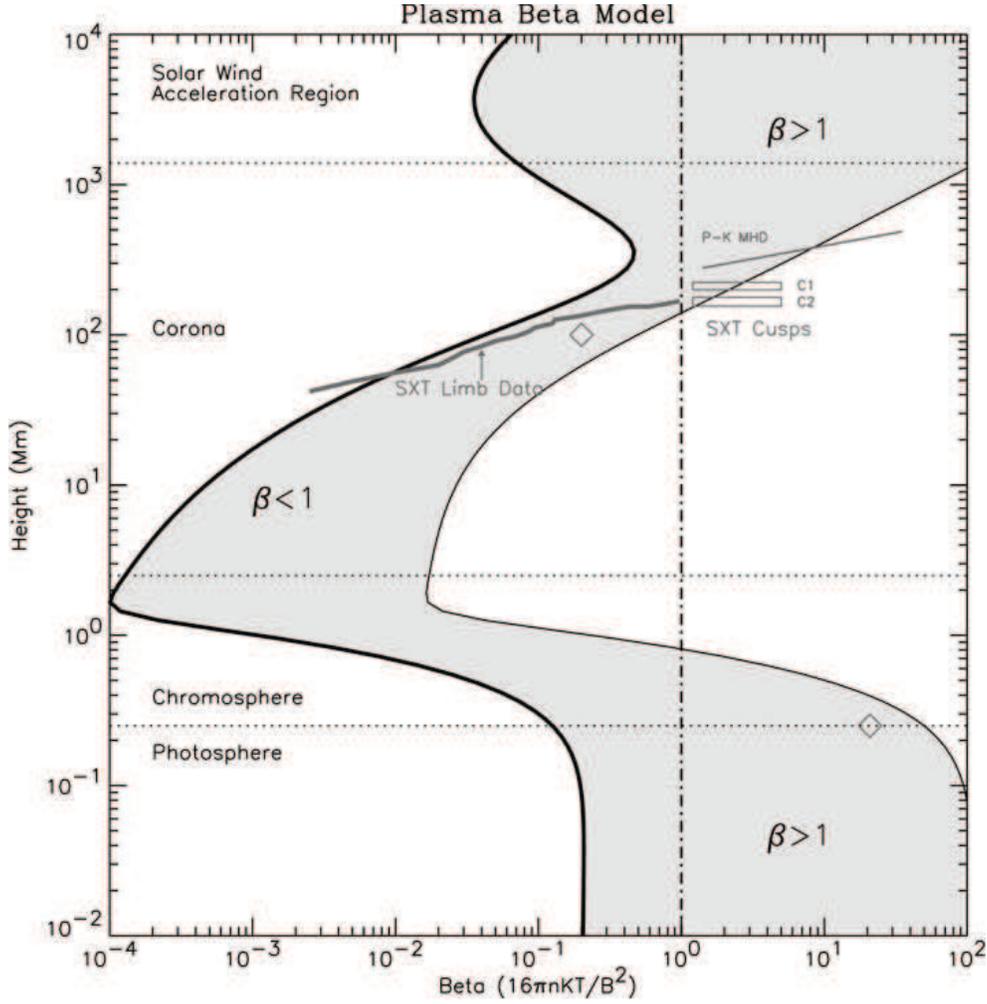}
  \caption{Plasma $\beta$ model over various active regions. The shaded area shows possible plasma $\beta$ corresponds to a sunspot region with 2500 G and a plage region with 150 G. Reproduced with permission from \cite{g01} Copyright 2001, Springer.}
  \label{fig:beta}
\end{figure*}


Gaining insight into the magnetic fields and plasma in the solar atmosphere is of great importance for studying various solar activities. So far magnetic field measurements in the photosphere are the most reliable. The chromospheric magnetic field measurements confront with much bigger uncertainties and it is difficult to determine the height of the detected magnetic field. The situation in the corona is ever worse since the corona is optical thin and magnetic field measurements have a line of sight integrated character. For constructing the three-dimensional magnetic field configurations in the chromosphere and corona, one may solve a set of equilibrium equations of magnetohydrodynamic (MHD) from given boundary data in the photosphere. This is the so-called magnetic field extrapolation. A popular and powerful extrapolation technique is the nonlinear force-free field (NLFFF) extrapolation which is designed to solve the following equations:
\begin{eqnarray}
\nabla\times \textbf{B} &=& \alpha(\textbf{r}) \textbf{B},\label{eq:ff}\\
\nabla\cdot\textbf{B} &=& 0,\label{eq:divb0}
\end{eqnarray}
where $\textbf{B}$ is the magnetic field and $\alpha(\textbf{r})$ is the force-free factor which is constant along a given field line. Equation (\ref{eq:ff}) indicates that the electric current is parallel to the magnetic field, which means a vanishing Lorentz force. Equation (\ref{eq:divb0}) is the solenoidal condition of the magnetic field. For recent reviews on the NLFFF extrapolation we refer to \cite{r13}, \cite{gcd17}, and \cite{ws21}.

The force-free equation (\ref{eq:ff}) assumes that the plasma $\beta$, defined as the ratio between the plasma pressure and magnetic pressure ($8\pi p/B^2$), of the solar atmosphere is low and all non-magnetic forces can be neglected. This is well justified in the most parts of the solar corona. However, it is not the case in the photosphere and lower chromosphere where the plasma pressure is comparable with the magnetic pressure \citep{g01}. Figure \ref{fig:beta} shows that plasma $\beta$ is about unit or larger roughly below 1 Mm above the photosphere. Non-magnetic forces are needed in these regions to balance the Lorentz force. Still assuming a static state, the equilibrium of the lower layers can be described by the magnetohydrostatic (MHS) equations:
\begin{eqnarray}
\frac{1}{4\pi}(\nabla \times \textbf{B})\times \textbf{B}-\nabla p - \rho \nabla \Phi & = & 0, \label{eq:force_balance}\\
\nabla \cdot \textbf{B} & = & 0 \label{eq:divb}.
\end{eqnarray}
In equation (\ref{eq:force_balance}), $p$ and $\rho$ are the plasma pressure and density, $\Phi$ is the gravitational potential. To close the above MHS system, we may relate pressure and density with temperature via an equation of state and prescribe an energy-balance equation as follows:
\begin{eqnarray}
p = \rho k_{B} T/m, \label{eq:ideal_gas}\\
\nabla\cdot(\textbf{K}\cdot\nabla T) + Q_{rad} + Q_{heat} = 0,\label{eq:mhs_energy}
\end{eqnarray}
with $T$ is the plasma temperature, $k_{B}$ the Boltzmann constant, $m$ the mean molecular weight of the gas, $Q_{rad}$ the source term accounts for heating or cooling owing to radiation, and $Q_{heat}$ the heat source owing to nano-flare, wave dissipation, or other heating mechanisms. $\nabla\cdot(\textbf{K}\cdot\nabla T)$ describes thermal conduction, where {\bf K} is the thermal conduction coefficient tensor. In particular, the cross-field thermal conduction is much less efficient than the field aligned component. With $Q_{rad}$ and $Q_{heat}$ being prescribed, the above set of partial differential equations (PDEs) (\ref{eq:force_balance})--(\ref{eq:mhs_energy}) can be solved, in principle, by imposing a set of boundary conditions ({\bf B}, $p$, and $\rho$) in a finite domain. The technique is called the MHS extrapolation. Note that there are a large number of works on the MHS equilibria for magnetosphere and Tokamak without considering gravity \citep[e.g.][]{s58,b04,hb93}. Here we only concentrate on the three-dimensional MHS equilibria for the solar case, in which gravity cannot be neglected. To our knowledge no review covering this field exists.

Compared with the established NLFFF extrapolation, the MHS extrapolation is much less developed mainly due to the insufficient spatial resolution of the routinely measured vector magnetogram in the past. For example, the spatial resolution of SDO/HMI is about 720 km \citep{ssb12} which is roughly 1/2 of the thickness of the non-force-free layer. Therefore an extrapolation based on the HMI data does not have sufficient spatial resolution vertically to study the thin non-force-free layer. Furthermore, the pressure scale height on the photosphere is about 150 km, which makes usage of HMI data in resolving this scale questionable. The situation, however, has been changed significantly by new ground-based instruments such as Daniel K. Inouye Solar Telescope (DKIST) which will carry out routine measurements of the magnetic fields in multiple layers with very high spatial resolution (0.03$''$ or 25 km on the solar surface at 500 nm) \citep{krw11}.

The aim of this review is to show both analytical and numerical extrapolation techniques which have been developed to solve the MHS equations when the photospheric vector magnetograms are used as the boundary input. The review is organized as follows. The analytical solutions of the MHS equations are discussed in Sect.~\ref{se:analytical}. An overview of the numerical extrapolation techniques are given in Sect.~\ref{se:numerical}. In Sect.~\ref{se:preprocess}, we describe the necessary boundary conditions for the MHS equilibrium and give an introduction of a preprocessing algorithm to deal with the inconsistency between the boundary and the MHS assumption. Finally, we summarize and discuss the challenges in the future development of the MHS extrapolation in Sect.~\ref{se:conclusion}.

\section{Analytical MHS solutions in 3D}\label{se:analytical}

The full set of the MHS equations (\ref{eq:force_balance})--(\ref{eq:mhs_energy}) has two aspects of equilibrium with the force equilibrium achieved by the magnetic fields and plasma, and thermodynamic equilibrium maintained by various energy processes of the plasma. The solution depends on how the two aspects of equilibrium are coupled. An analytical and complete treatment of this problem in 3D is not tractable. Most works has concentrated on seeking solutions of equations (\ref{eq:force_balance}) and (\ref{eq:divb}) without treating the energy equation self-consistently. The main interest is to find states that are admissible under force equilibrium. In spite of simplifying the equations greatly, they are by no means easy to be solved.

The MHS equilibrium requires some degree of symmetry in the magnetic field, which stems from the balance between Lorentz force and plasma forces \citep{p72}. The Lorentz force is highly anisotropic while the the plasma forces (pressure gradient and gravitational force) can be derived from isotropic scalar potential. By eliminating pressure term in the x and y components of equation (\ref{eq:force_balance}) with
\begin{equation} \label{eq:single-valued}
\frac{\partial^2}{\partial x\partial y}(p+B^2/8\pi)=\frac{\partial^2}{\partial y\partial x}(p+B^2/8\pi),
\end{equation}
we obtain the symmetry mathematically as follows
\begin{equation} \label{eq:compatity}
\frac{\partial}{\partial x}[(\textbf{B}\cdot\nabla)B_{y}] = \frac{\partial}{\partial y}[(\textbf{B}\cdot\nabla)B_{x}].
\end{equation}
Equation (\ref{eq:compatity}) is also called the compatibility relation which is a necessary but not sufficient condition for an MHS equilibrium. The compatibility relation is trivially satisfied for a system with an ignorable coordinate. For example, consider a two-dimensional plasma with $\frac{\partial}{\partial x}=0$, the left hand side of equation (\ref{eq:compatity}) then vanishes. The right hand side of equation (\ref{eq:compatity}) also vanishes because $(\textbf{B}\cdot\nabla)B_{x}=0$ which represents a null tension force in x-direction. The compatibility relation is also satisfied under rotational and helical symmetries. Substantial efforts have been made to treat the MHS equations by ignoring a coordinate \citep[e.g.][]{d53,l75,hhz81,ul81,aa89}. The symmetry introduced by the ignored coordinate render the MHS problem tractable.

However, the real solar structures observed are highly inhomogeneous and seldom show a high degree of geometric symmetry. On the other hand, the need of symmetry in the magnetic field does not necessarily mean an ignorable coordinate. Works by giving up the assumption of an ignorable coordinate have been conducted and some families of particular three-dimensional MHS solutions have been found. \cite{l82} found a family of three-dimensional MHS equilibria by assuming that the magnetic field lines lie in parallel vertical planes. Except for this assumption, structural variation of the system is allowed in all three dimensions. The constraint of the laminar magnetic field structure was relaxed in Low by demanding the magnetic tension force being vertical \citep{l84}:
\begin{eqnarray}
(\textbf{B}\cdot\nabla)B_{x} = 0, \label{eq:vertical_tensionx}\\
(\textbf{B}\cdot\nabla)B_{y} = 0. \label{eq:vertical_tensiony}
\end{eqnarray}
With equations (\ref{eq:vertical_tensionx}) and (\ref{eq:vertical_tensiony}), the compatibility condition is satisfied trivially. \cite{hhl83} built a slightly non-axisymmetric equilibrium with a perturbation theory. In their approach, an perturbation expansion was carried out to the originally axisymmetric magnetic flux tube.

Although these solutions are three-dimensional, they are not consistent with the measured line-of-sight (LOS) magnetogram due to some regularity imposed through the compatibility condition. It means that the previous solutions cannot be used to model the real Sun. This situation has not been changed until \cite{l85} (hereinafter L85) found a new class of three-dimensional MHS equilibria. The new equilibria which assume a horizontal electric current everywhere are capable of cooperating with arbitrary distribution of LOS magnetic field as the boundary condition. This characteristic makes the new solution applicable to model the real solar atmosphere, while previously only potential field \citep{s64} and linear force-free field \citep{ch77} had reached the level of sophistication. After that, a number of works have been stimulated based on the assumption made in L85. The following reviews these works.

\subsection{Analytical solution with untwisted magnetic field}

L85 derives a set of analytical and fully three-dimensional solutions of the MHS equations which are applicable to extrapolate the magnetic field and plasma from the observed magnetogram. The key assumption is to take the electric current to be everywhere perpendicular to the gravitational force. It follows that the transverse components of equation (\ref{eq:force_balance}) resolve into
\begin{equation}\label{eq:compatity2}
\frac{\partial B_z}{\partial x}\frac{\partial}{\partial y}\left(\frac{\partial\phi}{\partial z}\right)-\frac{\partial B_z}{\partial y}\frac{\partial}{\partial x}\left(\frac{\partial\phi}{\partial z}\right)=0,
\end{equation}
where $\phi$ is defined as
\begin{equation}\label{eq:b_general}
\textbf{B}=\left\{\frac{\partial \phi}{\partial x},\frac{\partial \phi}{\partial y}, B_{z}\right\}.
\end{equation}
Note that equation (\ref{eq:compatity2}) is equivalent of the compatibility relation equation (\ref{eq:compatity}). The general solution of equation (\ref{eq:compatity2}) is
\begin{equation}\label{eq:Psi}
B_z=\Psi\left(\frac{\partial\phi}{\partial z},z\right),
\end{equation}
where $\Psi$ is an arbitrary function of its arguments. Substitute equation (\ref{eq:Psi}) into magnetic divergence free equation (\ref{eq:divb}), then we obtain
\begin{equation}\label{eq:phi}
\frac{\partial^2\phi}{\partial x^2}+\frac{\partial^2\phi}{\partial y^2}+\frac{\partial}{\partial z}\Psi\left(\frac{\partial\phi}{\partial z},z\right)=0.
\end{equation}
For each prescription of $\Psi$, equation (\ref{eq:phi}) can be solved with the classical Neumann boundary condition of $\phi$
\begin{equation}\label{eq:bc}
S_0: \frac{\partial\phi}{\partial z}=\Psi^*(B_z,0),
\end{equation}
where $S_0$ is the bottom boundary and $\Psi^*$ is the inverse of $\Psi$ to express $\frac{\partial\phi}{\partial z}$ as a function of $B_z$ and $z$. Once $\phi$ is known, equation (\ref{eq:Psi}) determines $Bz$, and equation (\ref{eq:b_general}) yields {\bf B}. For each {\bf B} constructed, we can derive $p$ and $\rho$ directly through integrating the MHS equation (\ref{eq:force_balance}) at different directions.

The basic principle of the above solution is to reduce the MHS equations (\ref{eq:force_balance}) and (\ref{eq:divb}) to a single, scalar PDE (\ref{eq:phi}). The reduction is, of course, allowed by assuming a special type of electric current which is everywhere perpendicular to gravity and distributed smoothly in space. It is worth noting that the energy equation is omitted and the equation of state is not fixed \textit{a priori} in this method. The governing equations are thus closed by prescribing the function $\Psi$ of its arguments $\frac{\partial\phi}{\partial z}$ and $z$.

L85 studied a particular solution with
\begin{equation}
\Psi=\gamma\frac{\partial\phi}{\partial z}
\end{equation}
where $\gamma$ is a constant. They found magnetic field of the solution ($\gamma\ne 1$) can be obtained from the corresponding potential field ($\gamma=1$) by a uniform vertical expansion or contraction. Thus, the particular solution has a magnetic topology identical to that of the potential field. The same solution has also been expressed in spherical geometry with a point mass by \cite{bl86} (hereinafter BL86).

A more general formulation of the MHS equations was presented in \cite{l91} (hereinafter L91) by using electric current represented in terms of a pair of Euler potential $\mu$ and $\nu$:
\begin{equation}\label{eq:euler1}
\nabla\times\textbf{B}=\nabla\mu\times\nabla\nu.
\end{equation}
L91 assumed $\nu$ coincides with $\Phi$ (with $\Phi$ the gravitational potential) which sets the electric current to be perpendicular to gravity.
He found $\mu$ is a function of only two independent variables $\textbf{B}\cdot\nabla\Phi$ and $\Phi$:
\begin{equation}\label{eq:mu}
\mu=F(\textbf{B}\cdot\nabla\Phi,\Phi).
\end{equation}
It follows
\begin{equation}\label{eq:euler2}
\nabla\times\textbf{B}=\nabla F(\textbf{B}\cdot\nabla\Phi,\Phi)\times\nabla\Phi.
\end{equation}
The formulation (\ref{eq:euler2}) can be easily applied to any potential $\Phi$ which represents combination of any gravitational force and inertial force such as centrifugal force. Another advantage is to allow the magnetic field to have net twist, which will be discussed in the following subsection.

The MHS equations have also been reduced by using Clebsch representation, which have been presented by \cite{o85a,o85b}. To obtain an analytical solution, \cite{o85a} also used a special type of electric flow which is perpendicular to the gravity.

\subsection{Analytical solution with twisted magnetic field}

The main disadvantage of the case studied above is that the magnetic field does not have a net twist. The only exception exists when the magnetic field lies in the surfaces of constant $\Phi$, which is way far from reality.

To circumvent the deficiency, L91 added a field-aligned electric current component. Then the two current systems can be expressed, in Cartesian coordinates with $\Phi=gz$, as
\begin{equation}\label{eq:jtF}
\nabla\times\textbf{B}=\alpha\textbf{B}+\nabla F(B_z,z)\times\hat{\textbf{z}},
\end{equation}
where $\alpha$ is a constant in space. Note that equations (\ref{eq:euler2}) and (\ref{eq:jtF}) share the same mathematic form of the current perpendicular to the gravity. With a prescribed $F$, the governing equations (\ref{eq:jtF}) and (\ref{eq:divb}) for four unknowns are in general complete as a system.

For the case that function $F$ is linear in $B_z$, e.g.,
\begin{equation}\label{eq:F}
F=f(z)B_z
\end{equation}
as suggested by L91, equation (\ref{eq:jtF}) can be reduced to a linear form
\begin{equation}\label{eq:jtF2}
\nabla\times\textbf{B}=\alpha\textbf{B}+f(z)\nabla B_z\times\hat{\textbf{z}}.
\end{equation}
Fourier transforming of equation (\ref{eq:jtF2}) with respect to $(x,y)\mapsto(h,k)$ (so $\textbf{B}(x,y,z)\mapsto\tilde{\textbf{B}}(h,k,z)$) leads to a single, second order, and ordinary differential equation of $\tilde{B_z}$ as
\begin{equation}\label{eq:fourierjt}
\frac{d^2\tilde{B_z}}{dz^2}+\left\{\alpha^2+(h^2+k^2)\left[f(z)-1\right]\right\}\tilde{B_z}=0.
\end{equation}
Then $\tilde{B_x}$ and $\tilde{B_y}$ can be expressed in terms of $\tilde{B_z}$ as
\begin{eqnarray}
\tilde{B_x}&=&\frac{i}{h^2+k^2}\left(h\frac{d\tilde{B_z}}{dz}+k\alpha\tilde{B_z}\right), \label{eq:tildeBx}\\
\tilde{B_y}&=&\frac{i}{h^2+k^2}\left(k\frac{d\tilde{B_z}}{dz}-h\alpha\tilde{B_z}\right). \label{eq:tildeBy}
\end{eqnarray}
An inverse Fourier transformation is thus applied to recover {\bf B}. It follows the distribution of plasma pressure and density:
\begin{eqnarray}
p&=&p_0(z)-\frac{1}{8\pi}f(z)B_z^2, \label{eq:mhsp}\\
\rho&=&-\frac{1}{g}\frac{dp_0}{dz}+\frac{1}{4\pi g}\left[\frac{1}{2}\frac{df}{dz}B_z^2+f\textbf{B}\cdot\nabla B_z\right]. \label{eq:mhsd}
\end{eqnarray}
The first terms of equations (\ref{eq:mhsp}) and (\ref{eq:mhsd}) describe the stratified atmosphere and the second terms are the modifications by the non-force-free magnetic field.

Low (hereinafter L92) assumed that $f(z)$ has the following structure
\begin{equation}\label{eq:fz0}
f(z)=ae^{-\kappa z},
\end{equation}
where $a$ and $\kappa$ control the intensity and characteristic length of Lorentz force \citep{l92}. Then the general solution of equation (\ref{eq:fourierjt}) takes the form:
\begin{equation}\label{eq:bz_fourier}
\tilde{B_z}=M_{h,k}J_s[qe^{(-\kappa z)}]+N_{h,k}J_{-s}[qe^{(-\kappa z)}],
\end{equation}
where $\tilde{B_z}$ represents the Fourier transformation associated with component $e^{i(hx+ky)}$, $J_s$ and $J_{-s}$ are the Bessel functions, $M_{h,k}$ and $N_{h,k}$ are integration constants, $q^2=4a(h^2+k^2)/\kappa^2$, and $s^2=4(h^2+k^2-\alpha^2)/\kappa^2$. In solar applications, $J_{-s}$ term is rejected by setting $N=0$ since the magnetic field vanishes at infinity high above photosphere. Then $M$ can be determined using the observed photospheric magnetogram. L92 applied the MHS solution to an artificial active region magnetogram and showed a lot of interesting features, such as density enhancement along the neutral line at the lower atmosphere, twisted magnetic flux rope, etc. The high density plasma above the neutral line supported by the helical magnetic flux rope is interpreted as chromospheric filament.

With alternative structures of $f(z)$ many families of the MHS solutions have been found. \cite{n95} reduced equation (\ref{eq:jtF2}) to a Schr\"{o}dinger type equation which has been intensively studied in quantum mechanisms. With the new tool he derived three families of the MHS solutions corresponding to three choices of $\xi(r)$ for a spherical self-gravitating body. Here
\begin{equation}\label{eq:xi}
r^2\xi(r)=1-\frac{1}{\eta(r)},
\end{equation}
where $\eta(r)$ the counterpart of $f(z)$ in Cartesian coordinates. \cite{pn00} also studied the MHS solutions for three different choices of $f(z)$ by the Green's function method. \cite{nw19} presented a solution with
\begin{equation}\label{eq:fz1}
f(z)=a\left[1-b\,\textmd{tanh}\left(\frac{z-z_0}{\triangle z}\right)\right],
\end{equation}
where \textit{a} and \textit{b} control the magnitude of \textit{f}, $\triangle z$ controls the width of a transition from non-force-free to force-free field. The height and width of the transition being specified by the model parameters makes the new solution more flexible, which is also important to keep the plasma pressure and density positive.

For the case that function $F$ is nonlinear in $B_z$, i.e.,
\begin{equation}\label{eq:F_nonlinear}
F=\frac{B_z}{4\pi}+f(z)B_z^n,
\end{equation}
\cite{n97} presented a family of solutions which have an arcade-like magnetic field topology.

An alternative mathematical formulation for calculating the MHS equilibrium was introduced by \cite{nr99} with the magnetic field represented in terms of poloidal and toroidal components
\begin{equation}\label{eq:pol_tor}
\textbf{B}=\nabla\times\nabla\times(P\hat{\textbf{z}})+\nabla\times(T\hat{\textbf{z}}),
\end{equation}
where P and T are two scalar functions. This has an advantage that equation (\ref{eq:jtF}) can be reduced to only one scalar PDE with the following form
\begin{eqnarray}
\nabla^2P+\alpha^2P+F(z,B_z)&=&0,\\
T-\alpha P&=&0.
\end{eqnarray}
Thus P gains full knowledge of the complete magnetic field, which is very similar to the well-known case of linear force-free magnetic field \cite{nr72}. Previously, however, the three components of the magnetic field have to be computed separately. The \cite{nr99} model has been calculated efficiently with a multi-grid approach implemented by \cite{mem13}.

Solutions of the MHS equations with more general external potentials, such as centrifugal force in a rotating system, have been described in both spherical coordinate \citep{l91,n09,an10} and cylindrical coordinate \citep{anr10,wn18}.

It is worth noting that, by taking divergence of equation (\ref{eq:jtF}), the two current systems are conserved separately \citep{l93}. An interesting inference from the uncoupled currents is that there exists magnetic flux-current surface where the magnetic flux surface and the current surface overlap \citep{cj13}. \cite{l93} showed that all MHS equilibria of symmetric system have uncoupled electric currents with two independent coordinates. He also pointed out that the coupled currents equilibria which require three independent coordinates are a distinct class of the MHS solutions from the uncoupled one.

\subsection{Determination of parameters $\alpha$, $a$, and $\kappa$}

\begin{figure*}
  \centering
  \includegraphics[width=15cm]{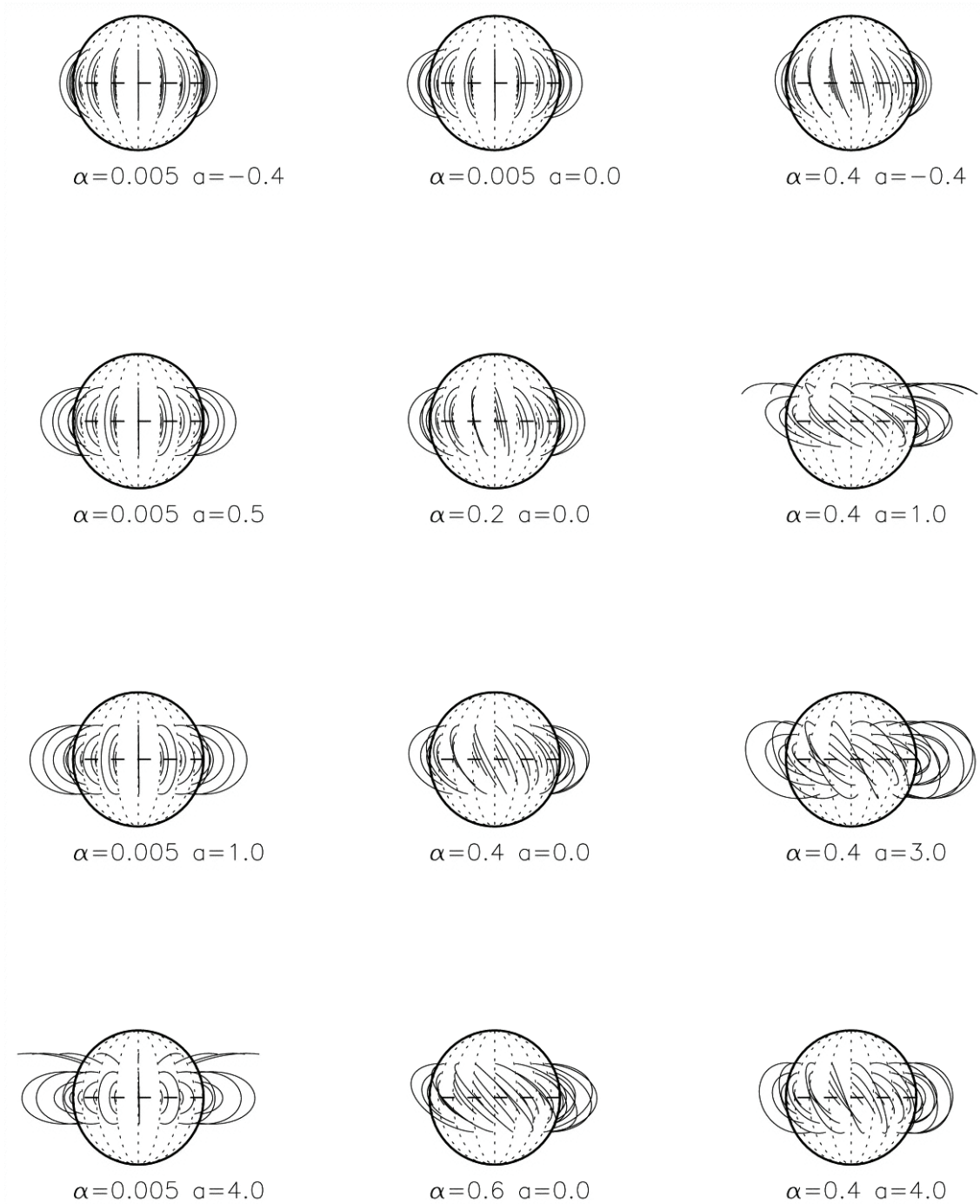}
  \caption{Influence of $\alpha$ and $a$ on the magnetic structure of the global corona. $\alpha$ and $a$ are normalized by $1/r_0$ and $r_0$, respectively, where $r_0$ is the radius of the Sun. It is worth to stress that definition of the parameter $a$ (see equation (1) in \cite{zhs00}) used in this figure is slightly different than dimensionless $a$ defined in equation (\ref{eq:fz0}). However, they both characterize the horizontal electric current system. The magnetic field line expands as $a$ increases, and it becomes twisted as $\alpha$ increases. Reproduced with permission from \cite{zhs00} Copyright 2000, AAS.}
  \label{fig:aalpha}
\end{figure*}

The L91 solution has three unknowns $\alpha$, a, and $\kappa$ (see equations (\ref{eq:jtF}) and (\ref{eq:fz0})). $\kappa$ is the reciprocal of the scale height above which the solution will become approximately force-free due to the low plasma $\beta$. The scale height often refers to the typical height of chromosphere where a significant interaction between the magnetic field and plasma takes place, leading to $1/\kappa=2$ Mm \citep{ads98,wnn15,wnn17}. The remaining parameters $\alpha$ and $a$ characterize field-aligned current system and horizontal current system, respectively. They may be varied to best fit observations. Figure \ref{fig:aalpha} illustrates how the magnetic structure changes with various choices of $\alpha$ and $a$. Note that $\alpha=a=0$ recovers potential field, while only $a=0$ recovers linear force-free field. If only $\alpha=0$ we recover L85 and BL86 model.

In the presence of vector magnetogram of an active region, $\alpha$ can be constrained by horizontal photospheric magnetic field vector. \cite{wnn17} used SUNRISE/IMaX vector magnetograms \citep{srb17} to compute $\alpha$ following an approach developed by \cite{hs04} for linear force-free fields:
\begin{equation}\label{eq:alpha}
\alpha=\frac{\sum\left(\frac{\partial B_y}{\partial x}-\frac{\partial B_x}{\partial y}\right)sign(B_z)}{\sum |B_z|},
\end{equation}
where the summation is taken over pixels of the magnetogram. In the absence of the vector magnetogram or the inaccurate measurements of the vector magnetic field due to poor signal-to-noise-ratio such as in the quiet regions, $\alpha$ is a free parameter to some extent. An additional restriction on $\alpha$, as discussed by \cite{ad98} and \cite{ads98}, is that $\alpha$ has a maximum value of $\alpha_{max}=2\pi/L$, where $L$ is the horizontal length of the computational box. L91 and L92 also showed that all modes with wavenumbers $h$ and $k$ in the domain $s^2=4(h^2+k^2-\alpha^2)/\kappa^2<0$ (see general solution (\ref{eq:bz_fourier})) oscillate with a nonvanishing amplitude as $z\rightarrow\infty$, and thus these modes should be rejected. To allow the solution to be applied to an arbitrary magnetogram, all wavenumbers should be included. This is ensured by $\alpha<2\pi/L$ since a periodic computational box with horizontal length $L$ has a minimum wavenumber $2\pi/L$. The property encountered here in the linear MHS solution has also been encountered previously in the study of linear force-free field. It is not accidental that the two models share the property because the linear MHS field becomes the linear force-free field in the limit of $\kappa z\gg 1$.

To constrain $a$ is a more challenging task. The dimensionless parameter $a$ controls the horizontal current which includes both magnetic field aligned and perpendicular components. The latter component gives rise to Lorentz force. Recall that the Lorentz force can be calculated by integrating Maxwell stress tenser $T$ over the entire surface over the computational volume \citep{m69,m74}. In the solar applications, a normalized parameter $\epsilon$ has been suggested by \cite{wis06} to check force-freeness of the magnetogram. $\epsilon$ is defined as
\begin{equation}\label{eq:epsilon}
\epsilon=\frac{|\sum B_xB_z|+|\sum B_yB_z|+|\sum(B_x^2+B_y^2-B_z^2)|}{\sum(B_x^2+B_y^2+B_z^2)},
\end{equation}
where the summation is taken over pixels of the magnetogram, each term in numerator represents Lorentz force in each direction. Thus both $a$ and $\epsilon$ relate to the Lorentz force. \cite{wnn17} suggested a linear relationship between them, empirically, as
\begin{equation}\label{eq:aepsilon}
a=2\epsilon.
\end{equation}
By using equation (\ref{eq:aepsilon}), $a$ can be specified in the presence of vector magnetogram. In the absence of vetor magnetogram, \cite{ads98} pointed out that $a$ has a maximum value $a_{max}$ to make the solution physical. To see this, the MHS solution (\ref{eq:bz_fourier}) is reduced by rejecting $J_{-s}$ term in solar applications:
\begin{equation}\label{eq:bz_fourier1}
\tilde{B_z}=M_{h,k}J_s[qe^{(-\kappa z)}].
\end{equation}
$J_s$ changes its sign as its argument increases, which implies an oscillation behavior of the magnetic field amplitude. \cite{ads98} took this behavior unphysical. They further found a reaches its maximum when $z=0$ and, under the extreme case, determined $a_{max}=1$.

\subsection{Boundary conditions}

As input the analytical MHS models require normal magnetic field $B_n$ measured at the photosphere. In the presence of high-quality vector magnetogram, $\alpha$, as a constant, can be constrained meaningfully. For a more general MHS case with
\begin{equation}
\textbf{B}\cdot\nabla\alpha=0 \label{eq:bdota}
\end{equation}
rather than $\alpha$ being a constant everywhere, \cite{l05} found that the closed contours of constant $\alpha$ at the boundary must come in pairs of equal total enclosed boundary flux. The topological property relating $\alpha$ and $B_n$ through ``hyperbolic'' equation (\ref{eq:bdota}) was first pointed out by \cite{a89} for force-free fields and is still applicable for the MHS case with two-current system. Then \cite{l05} derived a complete set of boundary conditions which contain the arbitrarily prescribed $B_n$ and a restricted $\alpha$ distribution. $\alpha$ needs to satisfy
\begin{equation}
\alpha(x,y,0)=G(g(x,y,0)),
\end{equation}
where $G$ is an even function and $g(x,y,0)$ takes the values of equal and opposite total boundary flux enclosed by each of $\alpha$ contour.

\subsection{Applications of the analytical MHS solutions to measured magnetograms}

Despite the limitations of the transverse current assumption, the analytical solutions have been used to study solar MHS equilibria in various situations. Since only the linear solution has been applied so far, the analytical MHS solution is often named the linear MHS solution in practice.

Some studies have focused on the lower atmosphere. \cite{ads98} modeled the magnetic field configuration of an H$\alpha$ flare by using L91 solution. They found the presence of dense plasma in regions of dipped field lines, which is in agreement with dark elongated features in H$\alpha$. The model also revealed that the flare kernels are located at separatrices correlated to ``bald patches'' (BPs) where field lines are tangential to the photosphere. The extrapolated BPs by the analytical MHS solutions have also been found associated with arch filament system and surge in chromosphere and transition region brightening \citep{flm01,mds02}. Comparisons between the analytical MHS solution and linear force-free solution have showed that the plasma effect helps to improve the magnetic configuration with a better correspondence with observation \citep{adm99,das08}. \cite{wnn15} and \cite{wnn17}, using L91 solution with magnetograms measured by the magnetograph IMaX \citep{mda11} on board SUNRISE \citep{sbd10,srb17}, modeled the magnetic field as well as plasma of a quiet region and an active region, respectively. Thanks to the high resolution of the magnetogram ($\sim 100$ km), the modeling resolves the thin non-force-free layer (typically less than 2000 km) with tens of grid points. Using the extrapolation data by \cite{wnn17}, \cite{jrs17} found the majority of slender Ca II H fibrils saw in SUNRISE/SuFI instrument overlap magnetic field lines at a height of around 700-1000 km. These field lines show a canopy-like dome over quiet areas.

Other studies have concentrated on global modeling of the solar corona \citep[see review by][]{my12}. The most widely used method to model the corona globally is the potential field source surface (PFSS) model which assumes a free current between the photosphere and the source surface. However, the current-free assumption is oversimplified because we have routinely observed large scale plasma structures, such as helmet streamers and coronal loop arcades, in multiple wavelengths which show appreciable interaction between magnetic field and plasma. \cite{bg91}, \cite{gb95} and \cite{gbl96} showed that the BL86 solution may account for the large-scale non-sphericity of the corona. The authors also found it necessary to include current sheets at equator and around the helmet streamer in the model to match white light observation of the corona during solar minimum. \cite{zh93,zh94,zh95} developed a global model to map the observed photospheric magnetic field into the corona and interplanetary space. A spherical ``cusp surface'' was defined at which all helmet streamer cusp points located identically. Below the cusp surface (or inner corona), the authors built an MHS atmosphere with BL86 solution. Then they extended the BL86 solution to larger radii with current sheets model developed by \cite{s71}. Their model better reproduces corona streamers and helmet structures as well as observed interplanetary magnetic field near the Earth's orbit. The BL86 solution have also been applied to recover the coronal fine structures observed during total solar eclipse \citep{adg09,yac18}. The more general solution based on BL86 by adding field-aligned current component \citep{n95} has also been used for corona modeling globally \citep{zhs00,r01,rwi08}. \cite{zhs00} found the magnetic arcade computed with $\alpha=0$ agrees best with the well-developed polar crown soft X-ray arcade, which suggests that the well-developed soft X-ray arcade has no field-aligned currents. \cite{rwi08}, using the \cite{n95} model, found the large structure of the magnetic field configuration is described reasonably well.

\section{Numerical approaches for the MHS extrapolation}\label{se:numerical}

Applying a linear MHS solution of equations (\ref{eq:force_balance}) and (\ref{eq:divb}) to model the sun has its limitation. The limitation has to do with the constant $\alpha$ and $a$ over the computational region, which makes the model fail to recover magnetic configurations with strong electric current concentration in part of the domain. A similar restriction occurs in the linear force-free modeling with globally constant $\alpha$. The linear MHS model also encounters a problem of negative plasma pressure in applications to solar quiet region. To circumvent these deficiencies, some numerical approaches have been developed recently to solve the MHS equation.

In the following, we briefly describe three different numerical approaches for computing solar MHS equilibria from photospheric vector magnetograms. The optimization method for MHS modeling was proposed by \cite{wn06} and further developed by \cite{zw18,zw19}. The MHD relaxation method for MHS modeling was developed by \cite{zwd13} and \cite{mki20}. The Grad-Rubin method for MHS modeling was proposed by \cite{gw13} and \cite{gbb16}. Both optimization method and MHD relaxation method use three components of vector magnetogram as magnetic boundary conditions, while the Grad-Rubin method uses vertical magnetic field and vertical electric current density as the boundary condition from observation. Every method requires the potential field as their very initial guess of the magnetic field.

\subsection{Optimization method}

We rewrite equation (\ref{eq:force_balance}) with a constant gravitational acceleration as follows:
\begin{equation}
(\nabla \times \textbf{B})\times \textbf{B}-\nabla p - \rho\hat{\textbf{z}} = 0. \label{eq:force_balance1}
\end{equation}
To display most clearly the mathematical structure of the equation, we have set relevant physical constants to unity. The optimization method for the MHS modeling is to find solution of the MHS equations (\ref{eq:force_balance1}) and (\ref{eq:divb}) by minimizing functional
\begin{equation}
L(\textbf{B},p,\rho)=\int_{V}B^{2}(\Omega_{a}^{2}+\Omega_{b}^{2})dV,\label{eq:L}
\end{equation}
where
\begin{eqnarray}
\textbf{$\Omega_a$} &=& B^{-2}[(\nabla \times \textbf{B})\times \textbf{B}-\nabla p - \rho\hat{\textbf{z}}], \\
\textbf{$\Omega_b$} &=& B^{-2}[(\nabla \cdot \textbf{B})\textbf{B}].
\end{eqnarray}
It is obvious that $L\ge0$. If the minimization process reaches $L=0$ then the MHS equations are fulfilled. Taking the functional derivative of the functional (\ref{eq:L}) with respect to the iteration parameter $t$ we have:
\begin{equation}
\frac{1}{2}\frac{dL}{dt}=-\int_{V}\left(\frac{\partial \textbf{B}}{\partial t}\cdot\textbf{F}-\frac{\partial p}{\partial t}\nabla\cdot\textbf{$\Omega_a$}+\frac{\partial\rho}{\partial t}\textbf{$\Omega_a$}\cdot\hat{\textbf{z}}\right)dV-\oint_{S}\left(\frac{\partial\textbf{B}}{\partial t}\cdot\textbf{G}+\frac{\partial p}{\partial t}\textbf{$\Omega_a$}\right)\cdot\textbf{dS},\label{eq:dL}
\end{equation}
where
\begin{eqnarray}
\textbf{F}&=&\nabla\times(\textbf{$\Omega_a$}\times\textbf{B})-\textbf{$\Omega_a$}\times(\nabla\times\textbf{B})+\nabla(\textbf{$\Omega_b$}\cdot\textbf{B})-\textbf{$\Omega_b$}(\nabla\cdot \textbf{B})+(\Omega_a^2+\Omega_b^2)\textbf{B},\\
\textbf{G}&=&\hat{\textbf{n}}\times(\textbf{$\Omega_a$}\times\textbf{B})-\hat{\textbf{n}}(\textbf{$\Omega_b$}\cdot\textbf{B}).
\end{eqnarray}

Given a fixed boundary condition the surface term vanishes. Then the gradient descent method is used to iterate {\bf B}, $p$, and $\rho$ by
\begin{eqnarray}
\frac{\partial\textbf{B}}{\partial t}&=&\mu_1\textbf{F},\\
\frac{\partial p}{\partial t}&=&-\mu_2\nabla\cdot\textbf{$\Omega_a$},\\
\frac{\partial\rho}{\partial t}&=&\mu_3\textbf{$\Omega_a$}\cdot\hat{\textbf{z}}.
\end{eqnarray}
With positive constants $\mu_1$, $\mu_2$, and $\mu_3$, $L$ decreases monotonically.

The optimization method was first proposed by \cite{wsr00} to calculate the NLFFF. \cite{wn06} extended the method by introducing plasma pressure in Cartesian geometry. They tested the code with an analytical MHS equilibrium (without gravity) similar to the force-free \cite{ll90} solution and found the reconstructed solution agrees well with the reference model. They also found the computing time of the MHS optimization is about 1000 times larger than the corresponding force-free optimization. The reason for the slow convergence is because plasma pressure in the reference model has a huge difference with 25 orders of magnitude.

\cite{wnr07} further extended the method by including gravitational forces in spherical geometry to calculate the global solar corona. The test of the code with a linear MHS solution by \cite{n95} showed the method works well and converges to the reference solution. However, the poles of the spherical geometry requires a sufficiently small time-step, which further slows down the optimization process. They suggested to use the Yin-Yang grid \citep{ks04} to address the problem.

Minimizing functional (\ref{eq:L}) subjects to constraints on plasma pressure and mass density which must be positive. However, it is not guaranteed by the previous methods. \cite{zw18} eliminated the constraints by using the variable transformation
\begin{eqnarray}
p&=&Q^2,\\
\rho&=&R^2.
\end{eqnarray}
Then the constrained problem is simplified to an unconstrained one
\begin{equation}\label{eq:minL}
minimize\quad L(\textbf{B},Q,R)=\int_{V}B^{2}(\Omega_{a}^{2}+\Omega_{b}^{2})dV,
\end{equation}
where
\begin{eqnarray}
\textbf{$\Omega_a$}&=&\frac{(\nabla\times\textbf{B})\times\textbf{B}-\nabla Q^2-R^2\hat{\textbf{z}}}{B^2+Q^2},\label{eq:oa1} \\
\textbf{$\Omega_b$}&=&\frac{(\nabla \cdot \textbf{B})\textbf{B}}{B^2+Q^2}.\label{eq:ob1}
\end{eqnarray}
Note that the denominators $(B^2+Q^2)$ in equations (\ref{eq:oa1}) and (\ref{eq:ob1}) are different than denominators $B^2$ in the previous method. Without plasma pressure in the denominator, $L$ will be dominated by regions with extremely weak magnetic field \citep{zw19}. The minimization problem (\ref{eq:minL}) then leads to
\begin{eqnarray}
\frac{\partial\textbf{B}}{\partial t}&=&\mu_1\textbf{F},\label{eq:itB}\\
\frac{\partial Q}{\partial t}&=&\mu_2Q\left[\lambda(\Omega_a^2+\Omega_b^2)-\nabla\cdot(\lambda\textbf{$\Omega_a$})\right],\label{eq:itQ}\\
\frac{\partial R}{\partial t}&=&\mu_3R\lambda\textbf{$\Omega_a$}\cdot\hat{\textbf{z}},\label{eq:itR}
\end{eqnarray}
where
\begin{eqnarray}
\textbf{F}&=&\nabla\times(\lambda\textbf{$\Omega_a$}\times\textbf{B})-\lambda\textbf{$\Omega_a$}\times(\nabla\times\textbf{B})+\nabla(\lambda\textbf{$\Omega_b$}\cdot\textbf{B})-\lambda\textbf{$\Omega_b$}(\nabla\cdot\textbf{B})+(1-2\lambda)(\Omega_a^2+\Omega_b^2)\textbf{B},\\
\lambda&=&\frac{B^2}{B^2+Q^2}.
\end{eqnarray}
The computational implementation involves the following steps.
\begin{enumerate}
  \item Calculate NLFFF \citep{w04} using vector magnetogram preprocessed with force-free criteria \citep{wis06}.
  \item Determine plasma pressure in the photosphere with formula $p=p_{quiet}-B_z^2/2$ with $p_{quiet}$ the plasma pressure in the quiet region. Distribute plasma along the magnetic field line by assuming gravitational stratification with a one-dimensional temperature profile.
  \item Iterate for $\textbf{B}, Q$ and $R$ by equations (\ref{eq:itB})-(\ref{eq:itR}). This step is repeated until $L$ reaches its minimum.
\end{enumerate}

Note that the bottom boundary condition of the plasma pressure ($p+B_z^2/2=p_{quiet}$) is simply the force balance condition of a hypothetically vertical magnetic flux tube. The rationality of the boundary setting can also be explained that the plasma pressure is related only with the vertical magnetic field in the linear MHS case. We take it as the lowest-order approximation in the nonlinear case.

The optimization code in the implementation of \cite{zw18} has recently been extended toward injecting the vector magnetogram slowly. By slow injection the initial magnetic field at the bottom boundary adjusted continuously by the gradient descent method over the total iterations of the calculation. The idea is to deal with measurement errors in photospheric vector magnetograms in particular in the transverse field. The slow injection approach is found to be capable of improving reconstructed magnetic field in both NLFFF and MHS calculations \citep{wti12,zw19}.

\cite{zw18,zw19} have tested the performance of the code applied to an analytical MHS solution (L91) and RMHD numerical simulation \citep{crc19}. It has been investigated how the unknown lateral and top boundaries, unknown bottom plasma boundary, initial conditions, and noise of magnetogram influence the solution. The code has also been applied to model the solar atmosphere with the high-resolution SUNRISE/IMaX vector magnetograms as well as moderate-resolution SDO/HMI vector magnetograms \citep{zws20,htc21,jwf21}.

\subsection{MHD relaxation method}

The MHD relaxation method uses time-dependent MHD equations to compute static equilibria. It has been widely applied to construct NLFFFs \citep{mbs88,r96,mjm97,vkk05,jf12,gxk16}. Recently, the method has been extended to construct MHS equilibria. The main idea is to solve either full or simplified MHD equations with plasma forces (e.g., pressure gradient and gravitational force) by using a friction term to relax the system into a quasi-static equilibrium.

\cite{zwd13} developed an MHD relaxation method by solving the full compressive MHD equations which are given as follows:
\begin{eqnarray}
\frac{\partial \rho}{\partial t} &=&-\nabla\cdot(\rho\textbf{v}),\\
\rho\left(\frac{\partial\textbf{v}}{\partial t}+\textbf{v}\cdot\nabla\textbf{v}\right) &=& \frac{1}{4\pi}(\nabla\times\textbf{B})\times\textbf{B}-\nabla p+\rho\textbf{g}-\mu\rho\textbf{v},\label{eq:momentum}\\
\frac{\partial e}{\partial t} &=& -p\nabla\cdot\textbf{v}-\mu\rho v^{2},\\
\frac{\partial {\textbf{B}}}{\partial t} &=& \nabla \times (\textbf{v}\times\textbf{B}),
\end{eqnarray}
where $-\mu\rho\textbf{v}$ is the friction term with coefficient $\mu$ represents the reciprocal of timescale of velocity damping. The initial condition is a potential magnetic field determined by the vertical component of the magnetogram, together with a gravity-stratified atmosphere including changes from the photosphere to the corona. Then the measured transverse magnetic fields are injected slowly at the bottom boundary \citep{r96}, while other physical quantities are fixed to their initial values. Deviation from equilibrium leads to plasma movements. Then the velocity is dissipated due to friction and finally a high-$\beta$ and non-force-free equilibrium is established in the lower atmosphere.

\cite{zwd13} also performed a test of the code on a numerical simulation of an active region emergence. Quantitative comparisons show that the method works well in extrapolating the non-force-free magnetic field. However, the distribution of plasma temperature is not well recovered. The code has also been applied to study solar activities in the lower atmosphere such as H$\alpha$ fibrils \citep{zwd16}, small-scale filament \citep{wlz16}, ultraviolet burst \citep{zsl17,tzp18,ctz19}, and blowout jet \citep{zwc17}. Other applications focus on large-scale structures of the magnetic fields in the corona also show the robustness of the method \citep{sgt18,mll18,jzs19,stz20,fhd20}.

Recently, \cite{mki20} developed an MHD relaxation method to reconstruct the non-force-free equilibrium by solving a set of simplified MHD equations given as follows:
\begin{eqnarray}
\frac{\partial\textbf{v}}{\partial t} &=& (\nabla\times\textbf{B})\times\textbf{B}-\nabla\tilde{p}-\frac{\tilde{p}}{H_0(z)}\hat{\textbf{z}}-\mu\textbf{v},\label{eq:momentum1}\\
\frac{\partial\textbf{B}}{\partial t} &=& \nabla\times(\textbf{v}\times\textbf{B}-\eta\nabla\times\textbf{B}),\\
\frac{\partial\tilde{p}}{\partial t} &=& -a^2\nabla\cdot\textbf{v},
\end{eqnarray}
where $-\mu\nu$ is the friction term,
\begin{equation}
\tilde{p}=p-p_0(z)
\end{equation}
is the pressure deviation from hydrostatic $p_0(z)$ of the background, and $H_0(z)$ is the pressure scale height defined by
\begin{equation}
H_0(z)\equiv\frac{T_0(z)}{g}.
\end{equation}
They assumed that both atmosphere and background atmosphere share the same temperature profile $T_0(z)$ which depends on $z$ only. Hence, the density deviation is given by
\begin{equation}
\tilde{\rho}=\frac{\tilde{p}}{T_0(z)},
\end{equation}
which removes $\tilde{\rho}$ in equation (\ref{eq:momentum1}). Note that both $\tilde{p}$ and $\tilde{\rho}$ can be negative. The initial condition is the same with that of \cite{zwd13}. Pressure deviation at the boundary during the relaxation process is extrapolated from the values in the computational area. The code has been tested in two-dimensional problem, but it is implemented in three dimensions.

\subsection{Grad-Rubin method}

The Grad-Rubin method was first proposed for fusion plasma by \cite{gr58}. Applications to NLFFFs can be found in \cite{s81}, \cite{abm99}, and \cite{w07}. The idea of the Grad-Rubin method is to replace the nonlinear equations to be solved by a system of linear equations, and then solve the linear equations iteratively until a fixed point is reached.

The first MHS implementation of the Grad-Rubin method was developed by Gilchrist et al. with $\textbf{g}=0$ \citep{gw13}. Gilchrist et al. extended the method to solve the MHS equations with $\textbf{g}\neq0$ \citep{gbb16}. In their approach, $p$ and $\rho$ are not independent since scale height $H(\textbf{r})$ is prescribed everywhere. To avoid instabilities due to spurious electric currents in weak-field regions, $p$ and $\rho$ are split into background ($p_0$ and $\rho_0$) and deviation ($\tilde{p}$ and $\tilde{\rho}$) components and each is computed separately. Then the MHS force-balance equation is reformulated in terms of $\tilde{p}$ as
\begin{equation}
\textbf{J}\times\textbf{B}-\nabla\tilde{p}-\left[\frac{p_0(z)+\tilde{p}}{H(\textbf{r})}-\frac{p_0(z)}{H_0(z)}\right]\hat{\textbf{z}}=0.\label{eq:force_balance2}
\end{equation}
The boundary condition on $\textbf{J}$, $\textbf{B}$, and $\tilde{p}$ are specified as follows:
\begin{eqnarray}
\textbf{B}\cdot\hat{\textbf{n}}|_{\partial V}&=&B_n,\label{eq:bnd_b}\\
\textbf{J}\cdot\hat{\textbf{n}}|_{\pm\partial V}&=&J_n,\label{eq:bnd_j}\\
\tilde{p}|_{\pm\partial V}&=&\tilde{p}_{bnd},\label{eq:bnd_p}
\end{eqnarray}
where $\pm\partial V$ is either $+\partial V$ where $B_n>0$ or $-\partial V$ where $B_n<0$. This means normal component of $\textbf{B}$ is prescribed over $\partial V$, while $\tilde{p}$ and normal component of $\textbf{J}$ are prescribed at only one polarity, not both.

The Grad-Rubin method decomposes the MHS equations (\ref{eq:force_balance2}) and (\ref{eq:divb}) into two hyperbolic parts for evolving $\tilde{p}$ and $\sigma$ (see definition below) along the magnetic field lines plus an elliptic equation to update the magnetic field by solving Amp\`{e}re's law. For iteration number $k$ one has to solve iteratively with the following steps:
\begin{enumerate}
  \item Calculate $\tilde{p}^{[k+1]}$ by solving
    \begin{equation}
    \nabla\tilde{p}^{[k+1]}\cdot\textbf{B}^{[k]}=\left[\frac{p_0(z)}{H_0(z)}-\frac{p_0(z)+\tilde{p}^{[k+1]}}{H(\textbf{r})}\right]B_z^{[k]}
    \end{equation}
    with the $\tilde{p}$ boundary condition (\ref{eq:bnd_p}).
  \item Calculate $\textbf{J}^{[k+1]}=\textbf{J}_\perp^{[k+1]}+\textbf{J}_\parallel^{[k+1]}$, where
    \begin{eqnarray}
    \textbf{$J_\perp$}^{[k+1]}&=&\textbf{B}^{[k]}\times\frac{\nabla\tilde{p}^{[k+1]}+\left[\frac{p_0(z)+\tilde{p}}{H(\textbf{r})}-\frac{p_0(z)}{H_0(z)}\right]\hat{\textbf{z}}}{\parallel\textbf{B}^{[k]}\parallel^2},\\
    \textbf{$J_\parallel$}^{[k+1]}&=&\sigma^{[k+1]}\textbf{B}^{[k]}.
    \end{eqnarray}
    Here $\sigma(\textbf{r})^{[k+1]}$, which is the counterpart of the force-free factor $\alpha$ in the force-free field, is a scalar function varies along the magnetic field line according to
    \begin{equation}
    \nabla\sigma^{[k+1]}\cdot\textbf{B}^{[k]}=-\nabla\cdot\textbf{J}_\perp^{[k+1]},
    \end{equation}
    which subject to boundary condition (\ref{eq:bnd_b}), (\ref{eq:bnd_j}), and
    \begin{equation}
    \sigma^{[k+1]}\bigg|_{\pm\partial V}=\frac{\textbf{J}_\perp^{[k+1]}\cdot\hat{\textbf{n}}-J_n}{B_n}\bigg|_{\pm\partial V}.
    \end{equation}
  \item Calculate $\textbf{B}^{[k+1]}$ by solving Amp\`{e}re's law and solenoidal condition
    \begin{eqnarray}
    \nabla\times\textbf{B}^{[k+1]}&=&\textbf{J}^{[k+1]},\\
    \nabla\cdot\textbf{B}^{[k+1]}&=&0,
    \end{eqnarray}
    with boundary condition (\ref{eq:bnd_b}).
\end{enumerate}
It is worth noting that all the above equations are linear in variables at step $k+1$ since variables at step $k$ are known. The iteration is initiated with a potential magnetic field embedded in a gravity-stratified atmosphere with a prescribed 1D temperature profile. To calculate $\textbf{B}^{[k+1]}$ in the third step, \cite{gbb16} uses a vector potential representation of the magnetic field. It turns out that the vector potential (in the Coulomb gauge) is a solution of Poisson's equation, which is solved effectively by multigrid scheme.

The code has been applied to a known analytical solution of the MHS sunspot model of \cite{l80}. Results show that the code converges quickly to a solution that closely matches the analytical solution and the discrepancy between the two solutions decreases with higher resolution. The run time has $\sim N^4$ scaling with $N$ the grid points size in each dimension, which is consistent with force-free codes of the Grad-Rubin type \citep{w06}. The Grad-Rubin code for MHS modeling is also found to be significantly slower than its counterpart for NLFFF. The main reason is that the MHS case has two field line tracing steps for $\sigma^{[k+1]}$ and $p^{[k+1]}$ while the NLFFF case only has one for $\alpha^{[k+1]}$.

\subsection{Other methods}

We also note a few other numerical methods that are being developed to solve the MHS equations.

The MHS equations with magnetic fields represented in terms of Euler potentials \citep{s70,s76}
\begin{equation}
\textbf{B}=\nabla\alpha\times\nabla\beta
\end{equation}
can be solved iteratively with Newton method \citep{zsb85,z87,pn94,rn99,rn02}. Since $\alpha$ and $\beta$ determine connectivity of the magnetic fields, this approach is very useful for studying evolution of the magnetic structures due to the photospheric plasma flow. However, for 3D problems there is no guarantee mathematically for a pair of Euler potentials to exist that can describe the magnetic field in the complete domain if the magnetic topology is complex, e.g. with magnetic null points and domains with different magnetic connectivity \citep{s70,s06}. Moreover, the Dirichlet boundary conditions for $\alpha$ and $\beta$ required in this approach are not possible to be derived from observation. Therefore, it would be difficult to use this method to calculate realistic MHS equilibria that could be used for solar magnetic field extrapolation.

Very recently, \cite{mfg21} proposed a novel numerical MHS model that solves directly for the force balanced magnetic field in the solar corona. This model is constructed with Radial Basis Function Finite Differences with 3D polyharmonic splines plus polynomials as the core discretization. It has been applied to reconstruct accurately the \cite{gl98} magnetic field configuration with proper boundary conditions and full information about the plasma forcing.

\section{How to deal with non-MHS boundaries?}\label{se:preprocess}

In the force-free case, several integral relations of the magnetic field at the boundary, called Aly's criteria, have to be fulfilled \citep{m69,m74,a84,a89}. However, studies have shown that these conditions are not always satisfied with the measured magnetic field in the photosphere \citep{mjm95,mcy02,t12,lsz13}. As a consequence, the non-force-free photospheric magnetogram has a negative impact on the extrapolation of the force-free field \citep{mds08}. \cite{wis06} has developed a preprocessing algorithm to use the Aly's criteria to derive a suitable boundary condition for the NLFFF extrapolation.

The same situation applies to the MHS case. In the following we will discuss the consistency check of the vector magnetogram for the MHS equilibrium and what we can do if the measured magnetic field is not compatible with the MHS assumption.

\subsection{Criteria for MHS Boundary}

The \textit{a priori} constraint is, when taking an isolated active region as the field of view (FOV), the magnetic field in the photosphere has to be flux-balanced
\begin{equation}
\int_SB_z(x,y,0)dxdy=0
\end{equation}
due to the solenoidal condition. Here $S$ is the photospheric surface.

\cite{zwi20} used the virial theorem to obtain a set of surface integrals as necessary criteria for an MHS equilibrium. Integrating equation (\ref{eq:force_balance1}) over an entire active region, note that the lateral and top surface integrals with magnetic terms approach zero as $\textbf{r}\rightarrow \infty$, we get net Lorentz force
\begin{align}
F_x &=\frac{1}{4\pi}\int_S B_xB_zdxdy = 0\label{eq:netforcex}, \\
F_y &=\frac{1}{4\pi}\int_S B_yB_zdxdy = 0\label{eq:netforcey}, \\
\begin{split}
F_z &=\frac{1}{8\pi}\int_S ({B_z}^2-{B_x}^2-{B_y}^2)dxdy\\
    &=\int_Spdxdy-g\int_V\rho dV \label{eq:netforcez}.
\end{split}
\end{align}
Equation (\ref{eq:netforcez}) implies that, in the MHS case, a non-zero $F_z$ is allowed. This is different from the force-free case in which each component of the net Lorentz force has to be zero. By a similar way with the cross product of equation (\ref{eq:force_balance1}) and {\bf r}, we obtain restrictions on the net torque of Lorentz force

\begin{align}
\begin{split}\label{eq:nettorquex}
T_x &= \frac{1}{8\pi}\int_S y({B_x}^2+{B_y}^2-{B_z}^2)dxdy\\
    &= -\int_Sypdxdy - g\int_V\rho ydV,\\
\end{split}
\\[1ex]
\begin{split}\label{eq:nettorquey}
T_y &= \frac{1}{8\pi}\int_S x({B_x}^2+{B_y}^2-{B_z}^2)dxdy\\
    &= -\int_Sxpdxdy - g\int_V\rho xdV,\\
\end{split}
\\[1ex]
\quad &T_z = \frac{1}{4\pi}\int_S (yB_zB_x-xB_zB_y)dxdy=0.\label{eq:nettorquez}
\end{align}
We see from equations (\ref{eq:nettorquex}) and (\ref{eq:nettorquey}) that the plasma may bring rotational moments in transverse directions. To summary, equations (\ref{eq:netforcex}) (\ref{eq:netforcey}) (\ref{eq:nettorquez}) are the necessary conditions which have to be fulfilled by the magnetogram in order to be suitable boundary for the MHS equilibrium.

Note that the restrictions on the boundary for the MHS equilibrium are fewer than those for the force-free field as, in the former case, non-vanishing $F_z$, $T_x$, and $T_y$ are allowed. Statistical studies have shown that there are considerable amount of magnetograms that do not satisfy Aly's criteria, which means these magnetograms are not suitable for the force-free field extrapolation. The statistics also reveal that it is $F_z$ that contributes most to the force in the magnetogram. Fortunately, $F_z$ has nothing to do with the constraints on the MHS boundary. Therefore, most of the non-force-free magnetograms can still be used for the MHS extrapolation directly. For the rest magnetograms which are not consistent even with the MHS equilibrium conditions, \cite{zwi20} proposed a preprocessing procedure to drive the measured data toward suitable boundary conditions.

\subsection{Preprocessing}

\cite{zwi20} proposed a preprocessing procedure by using integrals (\ref{eq:netforcex}) (\ref{eq:netforcey}) (\ref{eq:nettorquez}) to derive a more suitable boundary for the MHS extrapolation. The algorithm extended the preprocessing method that was developed by Wiegelmann et al. for the NLFFF extrapolation \citep{wis06}. To do so, we minimize the functional defined as follows:
\begin{align}
L  &=\mu_1L_1+\mu_2L_2+\mu_3L_3+\mu_4L_4,\label{eq:Lprep}\\
L_1&=\left(\sum_{p}B_xB_z\right)^2+\left(\sum_{p}B_yB_z\right)^2+\left(a_0-\sum_{p}({B_z}^2-{B_x}^2-{B_y}^2)\right)^2,\label{eq:L1}\\
\begin{split}\label{eq:L2}
L_2&=\left(a_1-\sum_{p}y({B_z}^2-{B_x}^2-{B_y}^2)\right)^2+\left(a_2-\sum_{p}x({B_z}^2-{B_x}^2-{B_y}^2)\right)^2\\
   &+\left(\sum_{p}(yB_xB_z-xB_yB_z)\right)^2,\\
\end{split}
\\[1ex]
L_3&=\sum_{p}(B_x-B_{xobs})^2+\sum_{p}(B_y-B_{yobs})^2+\sum_{p}(B_z-B_{zobs})^2,\label{eq:L3}\\
L_4&=\sum_{p}\left((\bigtriangleup B_x)^2+(\bigtriangleup B_y)^2+(\bigtriangleup B_z)^2\right),\label{eq:L4}
\end{align}
where
\begin{eqnarray}
a_0&=&\sum_{p}({B_{zobs}}^2-{B_{xobs}}^2-{B_{yobs}}^2), \\
a_1&=&\sum_{p}y({B_{zobs}}^2-{B_{xobs}}^2-{B_{yobs}}^2), \\
a_2&=&\sum_{p}x({B_{zobs}}^2-{B_{xobs}}^2-{B_{yobs}}^2),
\end{eqnarray}
are $F_z$, $T_x$, and $T_y$ of the original magnetogram, respectively. $L_1$ and $L_2$ correspond to the net force and net torque constraints. $L_3$ measures the deviation between the measured and preprocessed data. $L_4$ controls the smoothing, which is necessary for an optimization to obtain a good solution. Note that the introduction of $a_{i=0,1,2}$ differs from the MHS preprocessing from the NLFFF one by ensuring that the former procedure hardly changes $F_z$, $T_x$, and $T_y$. The aim of the preprocessing is to minimize $L$ in order to make all $L_n$ small. Then the boundary becomes suitable for MHS extrapolation and is as close as possible to the measured data at the same time. A method to specify the undetermined parameters $\mu_n$ has also been developed by \cite{zwi20}.

\section{Summary and Outlook}\label{se:conclusion}

In this review, we tried to give an overview of many different MHS extrapolation methods to model the solar atmosphere both analytically and numerically.

\begin{figure*}[b]
  \centering
  \includegraphics[width=15cm]{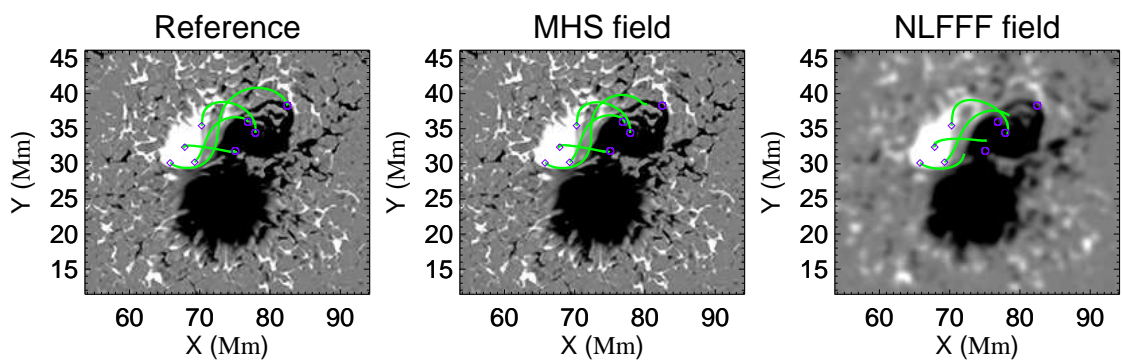}
  \caption{Selected field lines for test case in \cite{zw19}. The start points marked with rhombus are the same in each panel. Circles in each panel mark end points of the referenced field lines. Magnetogram of the NLFFF model is smoother than other two panels due to force-free preprocessing procedure \citep{wis06}. (a) Reference model \citep{crc19}, (b) MHS model implemented by \cite{zw18,zw19}, (c) NLFFF model implemented by \cite{w04}}.
  \label{fig:fld_line_compare}
\end{figure*}

The analytical MHS models have the advantage of leading to linear equations, which are computationally very cheap. However, one has to make a number of assumptions to derive the analytical solution, which limit the applicability of this kind of the MHS models to some extent. The numerical MHS models basically make no assumptions (except assuming a static state), and therefore are capable of modeling strong current concentrations locally which naturally leads to a nonlinear mathematical problem. However, the numerical MHS models are found to be really computationally expensive. \cite{gbb16} pointed out that more tracing steps (two in MHS case while only one in NLFFF case) and more stages involved in a single iteration slow down the MHS code significantly compared with the Grad-Rubin NLFFF code. The problem seems even worse in the optimization MHS code, which is 50 times slower than the optimization NLFFF code. The main reason is because there is a big difference in magnitude of the plasma parameters (e.g., pressure) through out the computational box and the code converges slowly. Recently, to reduce computational time, \cite{zw22} restricted the MHS extrapolation box to the non-force-free layer and then used the chromospheric vector magnetogram deduced from the MHS extrapolation to perform the NLFFF extrapolation. The combined approach reaches the accuracy of the MHS extrapolation and is moderately more efficient (still 7 times slower than the optimization NLFFF code).

The MHS models require high spatial resolution boundary conditions as input to resolve the fine structures in the solar lower atmosphere. While the spatial resolution of the observation is very high, the FOV usually limits to part of the observed active region (e.g., SUNRISE/IMaX covers a region of 37$\times$37 Mm with a pixel spacing of 40 km). To make the vector magnetogram to cover the entire active region which is necessary for a successful extrapolation by including more electric currents \citep{dsb09}, one has to either embed the high-resolution magnetogram into the lower one \citep{wnn17,zws20} or perform mosaic observations \citep{vdz21}. The combined magnetogram usually has a large number of grid points (e.g., up to 2000$\times$2000 even for a moderate active region with a FOV of 80$\times$80 Mm) which is also an challenge for extrapolations in terms of computing time and usage of computational resources. Implementation of MHS codes on massive parallel computer is needed currently or, at least, in the near future.

It is also worth noting that extrapolations with multiple observational data sets could improve the result dramatically. Methods that take advantage of combining magnetograms and EUV images have also been developed by \cite{mlm09,msd12} using a Quasi-Grad-Rubin method (QGR), \cite{a13,a16} based on vertical-current-approximation (VCA-NLFFF), and \cite{ciw15,cwi17} based on the optimization method (S-NLFFF). An issue that should be kept in mind when using multiple data sets is that providing all observations might overdetermines the boundary condition, and no solution can be obtained. Both QGR and VCA-NLFFF use normal component of the photospheric magnetic field, which rely on a well-posed boundary value problem. S-NLFFF uses full magnetic vector as boundary condition, which is overdetermined. However, S-NLFFF specifies measurement errors and, as a result, the optimal solution is allowed to deviate from observation within the errors. In the case of the MHS extrapolation, using multiple observations is also a promising way to get improved results. \cite{vdz21} investigated the similarity between the chromospheric magnetic fields inferred from direct inversion and the MHS extrapolation. They found the latter underestimates the transverse field strength as well as the amount of structure in the chromosphere. Therefore, more constraints (e.g., chromospheric magnetic field inversion at reliable sites) may be included in the MHS extrapolation in future developments, and it seems not too difficult for the optimization approach to do so by just adding additional terms in the functional to be minimized.

In numerical MHS models, the Grad-Rubin method uses well-defined boundary conditions made of $B_n$ in both polarities and $p$ and $J_n$ in one, which will not violate the ``hyperbolic'' transport properties of $p$ and field-aligned component of the current. However, both optimization method and MHD relaxation method require the full magnetic vector and $p$ in both polarities as input. They obviously overspecify boundary conditions if a $T(\textbf{r})$ distribution everywhere in the computational box is prescribed. However, the boundary issue can be somewhat relieved by allowing deviation of the boundary conditions from observation within measurement error.

So far, most of the efforts made regarding the MHS extrapolation concentrate on the model developments including testing and validating of the method with known reference models. Results, with these tests, have shown that the MHS extrapolation is capable of reconstructing field lines with higher accuracy (see figure \ref{fig:fld_line_compare}). However, applications of the MHS extrapolation to the observational data, from which we could learn new physics, are not as commonly seen as those of the NLFFF extrapolation. One of the reasons is that there are still some problems to be solved to make the MHS extrapolation as reliable as the NLFFF extrapolation in applying to solar data routinely. Another reason is that we do not have sufficient photospheric magnetic field measurements with sufficiently high spatial resolution ($\leq 100$ km). However, high-resolution measurements (e.g. with DKIST) are improving rapidly, which offer great opportunities to develop the MHS extrapolation. The MHS equilibrium also serves as a better initial condition than the NLFFF for the time-dependent MHD simulations.

\acknowledgments
XZ acknowledges the National Key R\&D Program of China (2021YFA1600500) and mobility program (M-0068) of the Sino-German Science Center. TN acknowledges financial support by the UKs Science and Technology Facilities Council (STFC) via Consolidated Grants ST/S000402/1 and ST/W001195/1. TW acknowledges financial support by DLR-grant 50 OC 2101.



\end{document}